\documentclass[%
 reprint,
showpacs,
 amsmath,amssymb,
 aps,
pra,
]{revtex4-1}

\usepackage[dvips]{graphicx}
\usepackage{dcolumn}
\usepackage[colorlinks=true,citecolor=blue,urlcolor=blue,linkcolor=blue,hyperindex]{hyperref}
\usepackage{bm}

\begin{document}


\title{Theory of Driven Nonequilibrium Critical Phenomena}

\author{Baoquan Feng}

\author{Shuai Yin}

\author{Fan Zhong}  \thanks{Corresponding author. E-mail: stszf@mail.sysu.edu.cn}
\affiliation{State Key Laboratory of Optoelectronic Materials and
Technologies, School of Physics and Engineering, Sun Yat-sen
University, Guangzhou 510275, People's Republic of China}

\date{\today}

\begin{abstract}
A system driven in the vicinity of its critical point by varying a relevant field in an arbitrary function of time is a generic system that possesses a long relaxation time compared with the driving time scale and thus represents a large class of nonequilibrium systems. For such a manifestly nonlinear nonequilibrium strongly fluctuating system, we show that there exists universal nonequilibrium critical behavior that is well described incredibly by its equilibrium critical properties. A dynamic renormalization-group theory is developed to account for the behavior. The weak driving may give rise to several time scales depending on its form and thus rich nonequilibrium phenomena of various regimes and their crossovers, negative susceptibilities, as well as violation of fluctuation-dissipation theorem. An initial condition that can be in either equilibrium or nonequilibrium but has longer correlations than the driving scales also results in a unique regime and complicates the situation. Implication of the results on measurement is also discussed. The theory may shed light on study of other nonequilibrium systems and even nonlinear science.
\end{abstract}

\maketitle


\section{\label{sec:level1}Introduction}
Although equilibrium statistical physics has achieved great success, equilibrium systems are exception rather than the rule: nonequilibrium phenomena are far more abundant and thus attract considerable attention~\cite{Schmittmann,Marro,Hinrichsen,Jiang,Racz,Ruelle,Odor,Mazenko,Rammer,Chou,Henkel,Klages}. Even though a systematic framework similar to the equilibrium statistical mechanics is still elusive, unifying principles for some nonequilibrium systems have emerged. For small systems, for instance, the key role of fluctuations and various fluctuation theorems for them have given birth to stochastic thermodynamics~\cite{Seifert} and thermodynamics of information~\cite{Parrondo}. But how about macroscopic systems?

Nonequilibrium systems are disparate without a systematic classification. One way is to classify them according to the driving that brings them into nonequilibrium states. One category is then to change some controlling parameters of a system instantaneously to their new values. The system then enters a nonequilibrium relaxation process. It can result in either a new equilibrium state or a nonequilibrium steady state~\cite{Schmittmann,Marro,Hinrichsen,Jiang,Racz,Ruelle,Chou,Henkel,Klages,Sasa} depending on whether a finite current flows through the system. Another one is to change the parameters infinitely slowly~\cite{Mazenko}. This is an adiabatical way that is usually invoked in theoretical studies such as linear responds to study small deviations from equilibrium. The third category, on which we focus here, is to change the parameters within a finite time. Jarzynski's work theorem for small systems was derived for such processes~\cite{Jarzynski}. For a macroscopic system, however, such a driving does not necessarily take it into nonequilibrium states.

Whether a driven system is in equilibrium or not depends on its relaxation time and the time scale of the driving. If the former is shorter than the latter, the system can follow the variation of the external driving adiabatically and hence stays in quasiequilibrium or adiabatic states. Only in the reverse case can a system fall genuinely out of equilibrium. The larger the difference between the two time scales, the more strongly the system deviates from equilibrium states. A system in a glassy state has a long relaxation time. A system close to its critical point also possesses a divergent correlation time. Moreover, the equilibrium properties of the latter system has a well-established theoretical framework of the renormalization-group (RG) theory~\cite{Wilson,Mask,Justin,Amit}. Accordingly, driving a system in the vicinity of its critical point within a finite time is a prototype of genuine nonequilibrium systems and is well fitted for studying whether universal nonequilibrium behavior exists or not. For comparison, relaxing a critical system of the first category has led to a critical initial slip~\cite{Janssen} and the corresponding method of short-time critical dynamics has been applied extensively to estimate critical properties~\cite{ZhengB,Ozeki,Albano}.

Indeed, some aspects of such driven systems have already been studied. On the one hand, the Kibble-Zurek (KZ) mechanism~\cite{Kibble1,Kibble2,Zurek1,Zurek2}, first proposed in cosmology and then applied to condensed matter physics, provides a mechanism for nonequilibrium topological defect formation after a system is cooled through a continuous phase transition to a symmetry-broken ordered phase. Upon combining the equilibrium scaling near the critical point with the adiabatic--impulse--adiabatic approximation, a universal KZ scaling for the defect density has been proposed~\cite{Zurek1,Zurek2,revqkz1,revqkz2}. It has then been tested intensively in many systems, ranging from classical \cite{Laguna1,Laguna2,Laguna3,Laguna4,Laguna5,Laguna6,Laguna7,Laguna8,KZtest12,Laguna9,Laguna10,Laguna11,Laguna12,Laguna13,Laguna14,Laguna15,Laguna15} to quantum~\cite{qkz1,qkz2,qkz3,qkz4,qkz5,qkz6,qkz7,qkz8,qkz9,inexper1,inexper2,inexper3,qkz10,revqkz1,revqkz2}. A recent experiment on the Bose-Einstein condensation found agreement with the KZ scaling~\cite{Laguna15}, though another one about the Mott insulator to superfluid transition on optical lattices concluded further theories were needed~\cite{qkz10}. This is in line with the fact that most experimental results require additional assumptions for interpretation of their consistency with the KZ scaling~\cite{inexper4}. As defect counting is not easy~\cite{Das} and whether phase ordering plays a role or not is yet to be clarified~\cite{Biroli}, it was proposed recently to detect the scaling of other observables~\cite{Das}. The two recent experiments, for example, measured the domain size~\cite{Laguna15} and the correlation length~\cite{qkz10} instead of the defect density.

On the other hand, finite-time scaling (FTS)~\cite{Zhong1,Zhong2} offers a different perspective on the problem. From the analogy between the space domain of a diverging correlation length that may get longer than a system's size and the time domain of a diverging correlation time $t_{\rm eq}$ that may be longer than its allowable relaxation time, FTS was proposed as a temporal counterpart of the well-known finite-size scaling (FSS). A linear driving with a rate $R_1$ was found to specify a readily tunable driving time scale $t_{R_1}$ that is asymptotically proportional to $R_1^{-z/r_1}$, where $r_1$ is the RG eigenvalue of $R_1$ and $z$ the dynamic critical exponent. Similar to FSS, in the FTS regime, $t_{R_1}<t_{\rm eq}$; the system falls out of equilibrium and just lies in the impulse regime of the KZ mechanism. This means that $t_{R_1}$ divides the adiabatic and impulse regimes and governs the evolution of the latter, thus improving its understanding~\cite{Huang1}. FTS has been successfully applied to classical~\cite{Zhong1,HuangX,Xiong1,Xiong2,Huang1,Zhong2} and quantum systems~\cite{Yin1,Yin3,Yin2} to determine their critical properties. In particular, a positive specific-heat critical exponent $\alpha$ and thus violation of the bound for the correlation-length critical exponent $\nu$~\cite{Chayes} was found for a randomness-rounded first-order phase transition~\cite{Xiong2}, corroborating by subsequent studies~\cite{Bellafard,Bellafard1}, and the critical behaviors of heating and cooling were observed to be qualitatively different~\cite{Huang1}. In addition, the initial slip has been recently combined with FTS, extending the KZ mechanism to beyond adiabaticity~\cite{Huang2}.

So far, most work for KZ mechanism and FTS considers primarily a linear driving across the critical point. For a driving that is not exactly linear in time $t$, it is linearized near the critical point \cite{revqkz1,qkz5,protocol}. For a nonlinear driving, a monomial form $t^n$ is usually considered with a non-unity constant $n$~\cite{qkz5,revqkz1,Zhong1,protocol,Sandvik}. An advantage of these forms is that essentially only one parameter is involved and the driving may appear simple. However, it is not easy to confirm a driving to be linear in experiments. Questions arisen naturally are then how about a general driving of arbitrary form within a finite time. How can one generalize the understanding gained in FTS to such a general case? Does such a case possess universal behavior, and if yes, how to describe it?

Note that driving in a general form within a finite time near a critical point is highly nontrivial. Within a finite time, the system inevitably falls out of equilibrium due to critical slowing down. Also, it is characterized by a set of usually non-unity critical exponents and thus behaves strongly nonlinearly there~\cite{Zhong06}. The theory of FTS~\cite{Zhong06,Zhong1,Zhong2}, which deals with both the nonequilibrium behavior in the FTS regime and the equilibrium behavior in the adiabatic regime, is mathematically a stochastic nonlinear time-dependent Landau-Ginzburg equation~\cite{Hohenberg,Folk,Tauber}. Upon a general driving, one of the controlling parameter becomes an arbitrary function of time. So, whether such a nonlinear partial differential equation in nonequilibrium situations shows universal behavior is surely not obvious even though the magnitude of the driving is small.

In this paper, we study the behavior of a system that is driven weakly close to its critical point within a finite time in a form that does not generate resonances but otherwise is arbitrary. We shall show that the system exhibits universal nonequilibrium critical behavior as may be expected. What is unexpected is that, incredibly, this driven critical behavior, far off equilibrium as the fluctuation-dissipation theorem is violated and the susceptibility can take on negative values, is well described by only the equilibrium static and dynamic critical exponents, though the scaling functions can still involve singularities that need the exponent of the driving. A dynamic RG theory will be developed for the system to account for its universal nonequilibrium critical behavior. It shows that there exist multiply time scales determined by the driving parameters themselves and their combination. As a result, the system can lie in different nonequilibrium regimes controlling by different time scales, with crossovers between them depending on the parameters. This generalizes the theory of FTS in which a single time scale arising from a linear driving governs the evolution of the system in the nonequilibrium regime. An initial condition that has longer correlations than the driving scales also gives rise to a unique regime and complicates the situation. This is opposite to the critical initial slip in which a nonequilibrium initial state has shorter correlations than those of the equilibrium state. Our theory furnishes a corrected understanding of experimental measurements in which an external driving is applied to a system with long relaxation times. As the system studied is a generic nonequilibrium one, the theory may shed light on the study of other nonequilibrium systems. It may also be instructive to nonlinear science as the driving may help to probe scaling behavior there.

We note that the driving form can be arbitrary except that sometimes the driving itself may generate some kinds of resonance depending on the systems considered. At present, we can only detect this from the results \emph{a posteriori}. In case they do not fit the theory, some resonance may be in effect.

In the following, we shall first develop a dynamic RG theory in Sec.~\ref{sec:RG} and study the effect of initial conditions in Sec.~\ref{sec:initial}. We then apply the theory to several specific forms of driving and discuss its implication to measurements in Secs.~\ref{sec:examples} and~\ref{sec:wArgue}, respectively. In order to test our result, we perform Monte Carlo (MC) simulations using the model and method in Sec.~\ref{sec:model} with the results being detailed in Sec.~\ref{sec:result}. Conclusions are given in Sec.~\ref{sec:conclusion}.

\section{\label{sec:RG}Dynamic RG theory}
In this section, a dynamic RG theory is first developed to analyze the universal behavior of a system driven by a general temporal form near its critical point. Then different timescales are identified and crossovers are briefly discussed from the scaling forms obtained. We only consider the cases in which the system starts with an equilibrium initial condition far away from the critical point. The effect of initial conditions is left to Sec.~\ref{sec:initial}.

\subsection{\label{drg}Theory}
The dynamic RG theory for a system with a driving was initiated in a theory of first-order phase transitions~\cite{zhongchen}. It was then applied back to critical phenomena~\cite{Zhong06,Zhong2}. Here we shall generalize the theory to a driving of a general form and identify the restriction on the driving with which different behavior may emerge.

Any relevant parameter such as the temperature $T$ or an externally applied field can serve as a driving field. Without loss of generality, we use the terminology of magnetism and choose the external magnetic field $H$ as the driving throughout. For clarity, we shall often set the reduced temperature $\tau=T-T_c=0$ and ignore the effect of finite system sizes $L$, where $T_c$ is the critical temperature. They can be taken into account straightforwardly, though finite-time finite-size scaling may emerge in cooling when $L$ is considered~\cite{Huang1}.

Consider a $\phi^4$ free energy functional~\cite{Mask,Justin,Amit}
\begin{equation}\label{eq:F}
\mathcal{F}[\phi]=\int{\mathrm{d}\mathbf{r}\left[\frac{1}{2}\bar{\tau}\phi^2+\frac{1}{4!}g\phi^4+\frac{1}{2}(\nabla \phi)^2-H\phi\right]},
\end{equation}
where $\phi$ is a coarse-grained field variable, $g$ a coupling constant, and $\bar{\tau}$ the distance to the mean-field $T_c$ at $H=0$. The dynamics is governed by the Langevin equation
\begin{equation}\label{eq:dye}
\frac{\partial \phi}{\partial t}=-\lambda \frac{\delta \mathcal{F}}{\delta \phi}+\zeta,
\end{equation}
where $\lambda$ is a kinetic coefficient and $\zeta$ is a Gaussian white noise satisfying $\langle \zeta(\mathbf{r},t) \rangle=0$ and $\langle \zeta(\mathbf{r},t) \zeta(\mathbf{r'},t') \rangle=2\lambda T\delta(\mathbf{r}-\mathbf{r'})\delta(t-t')$. The dynamic model Eqs.~(\ref{eq:F}) and (\ref{eq:dye}) constitutes the simplest equilibrium critical dynamics of Model A for the non-conserved order parameter~\cite{Hohenberg}. It is a nonlinear stochastic partial differential equation that cannot be solved exactly generally. Moreover, perturbation expansions near the critical point are plagued with infrared divergences~\cite{Mask}.

However, universal long-wavelength long-time properties can be found by the RG theory without solving the equation. This can be done systematically using field-theoretic methods~\cite{Justin,Amit}. It has been shown that the model Eqs.~(\ref{eq:F}) and (\ref{eq:dye}) is equivalent to a dynamical field theory described by the dynamical functional~\cite{Janssen79,janssen,Tauber}
\begin{equation}\label{eq:dyfunctional}
\begin{split}
I[\phi,\tilde{\phi}]=\int \mathrm{d}\mathbf{r}\mathrm{d}t&\left\{\tilde{\phi}\left[\dot{\phi}+\lambda\left(\bar{\tau}-\nabla^2\right)\phi \right.\right. \\
&\ \ \left.\left.+\frac{1}{3!}\lambda g\phi^3-\lambda H\right]-\lambda T\tilde{\phi}^2\right\},
\end{split}
\end{equation}
where $\tilde{\phi}$ is a response field~\cite{martin}. In the field-theoretic framework, the universal critical behavior are determined by the renormalization factors that remove the divergences arising at long times and when the underlying lattice constant of the original theory is set vanishing.

For a constant external field $H$ and an equilibrium initial condition, because of the supersymmetry of $I[\phi,\tilde{\phi}]$~\cite{Justin}, it is well known that only the following four independent renormalization factors $Z$ defined as
\begin{equation}\label{eq:RGZ}
\begin{array}{l l}
\phi \rightarrow\phi_0=Z_{\phi}^{1/2}\phi, & \tilde {\varphi }\rightarrow\tilde{\phi_0}=Z_{\tilde{\phi}}^{1/2}\tilde{\phi},\\
g\rightarrow g_0=N_d\mu^{\epsilon}Z_{\phi}^{-2}Z_u u, &\lambda \rightarrow \lambda_0=Z_{\phi}^{1/2}Z_{\tilde{\phi}}^{-1/2}\lambda,\\
\bar{\tau} \rightarrow \bar{\tau}_0=Z_{\phi}^{-1}Z_{\tau}\tau+\bar{\tau}_c, & H \rightarrow H_0=Z_{\phi}^{-1/2}H,
\end{array}
\end{equation}
are needed, where the subscripts $0$ denote bare parameters, $\mu$ is an arbitrary momentum scale, $\bar{\tau}_c$ the shifted critical point, $\epsilon=4-d$, and $N_d=2/[(4\pi)^{d/2}\Gamma(d/2)]$ with $d$ being the space dimensionality and $\Gamma$ the Euler Gamma function. We have directly renormalized the field $H$ in Eq.~(\ref{eq:RGZ}). It results from the expansion of response functions with $H$~\cite{Justin,Amit}. The four $Z$ determine the fixed point and three independent critical exponents including the dynamic one.
However, when the initial state is out of equilibrium with a short correlation in the vicinity of the critical point, it was found that another new $Z$ factor is required to cancel the new divergence due to the initial time. This leads to an independent critical initial-slip exponent~\cite{Janssen,janssen}.

Now, for a driving with a time-dependent $H$, upon ignoring the effects from the nonequilibrium initial conditions, which have been studied~\cite{Huang2}, whether new exponents are needed hinges on whether new intrinsic divergences are generated. Since we only focus on a spatially homogeneous driving, possible new divergences can only stem from the time domain. When $H$ blows up with $t$ as in the linearly driving case, a divergence at $t\to\infty$ always presents. But it is extrinsic as it arises from the driving itself. A nontrivial divergence must originate from a resonance-like interaction of the driving with the system considered. This must then result in new exponents. Interesting as it is, this is not the case on which we focus here as our aim here is to bring a system out of equilibrium. In this case, we can again expand the response function with $H$ at each instant as in the time-independent case. Therefore, the four $Z$ suffice to remove all the divergences and no new exponents are needed! We shall meet a new singularity in some monomial driving, but that is generated completely by the form of the field itself and no critical exponents are needed there.

With the renormalization factors, the universal behavior can be determined by the RG equation. It is~\cite{Zhong06,Zhong2}
\begin{equation}\label{eq:RGeq}
\left[\mu\partial_{\mu}+\varsigma\lambda\partial_{\lambda}+\beta\partial_{u}+ \gamma_{\tau}\tau\partial_{\tau}+\frac{1}{2}\gamma_{\phi} (H\partial_{H}+1)\right]M=0,
\end{equation}
where $M\equiv\langle\phi\rangle$ and the Wilson functions are defined as
\begin{equation}\label{eq:RGWilson}
\begin{array}{l l}
\varsigma=\mu\partial_{\mu}\ln\lambda,\qquad\qquad &\gamma_{\phi}=\mu\partial_{\mu}\ln Z_{\phi},\\ \gamma_{\tau}=\mu\partial_{\mu}\ln\tau,\qquad \qquad &\beta(u)=\mu\partial_{\mu}u.
\end{array}
\end{equation}
At the fixed point $u=u^*$ at which $\beta(u^*)=0$, combining the solution of Eq.~(\ref{eq:RGeq}) with the result of dimension analysis, one arrives at
\begin{equation}\label{eq:Mb}
M(t,H,\tau)=b^{-\beta/\nu}M(tb^{-z},Hb^{\beta\delta/\nu},\tau b^{1/\nu}).
\end{equation}
where $b$ is a length rescaling factor and the critical exponents are given as usual by
\begin{equation}\label{eq:sol2exp}
\begin{split}
\eta=\gamma_{\phi}^*,\qquad \nu^{-1}=2-\gamma^{*}_{\tau},\qquad z=2+\varsigma^*, \\
\beta/\nu=(d-2+\eta)/2,\ \ \delta=(d+2-\eta)/(d-2+\eta)
\end{split}
\end{equation}
with the stars marking the values at the fixed point. Equation~(\ref{eq:Mb}) gives the scale transform of $M$ and applies to both a constant and a time-dependent $H$. It can give rise to various scaling forms.
For example, choosing a scale such that $\tau b^{1/\nu}$ is a constant leads to
\begin{equation}\label{eq:Mtau}
M(t,H,\tau)=\tau^{\beta}f_{\tau}(t\tau^{\nu z},H \tau^{-\beta\delta}),
\end{equation}
where $f_{\tau}$ is a universal scaling function.

We now turn to the parameters that specify the time-dependence of the driving. Let $H=H(t, B_1,\dots,B_p)$, where $B_i,i=1,\dots,B_p$ are $p$ independent parameters. Using $B_i$ and $t$ instead of $H$ and $t$ as variables because they are not independent, we can write formally the RG equation as~\cite{Zhong2}
\begin{equation}\label{eq:RGeqNew}
\left[\varsigma t\partial_{t}+\sum_{i=1}^{p} \left(\gamma_{B_i}B_i\partial_{B_i}\right)+\mu\partial_{\mu}+\beta\partial_{u}+\frac{1}{2}\gamma_{\phi}\right]M=0.
\end{equation}
where $\gamma_{B_i}=\mu \partial_{\mu}\ln B_i$ are the Wilson function of $B_i$ and we have replaced $\lambda$ with $t$ directly and suppressed $\tau$ by considering the critical theory only. As a result, similar method then gives rise to
\begin{equation}\label{mbb}
M(t,B_1,\dots,B_p)=b^{-\beta/\nu}M(tb^{-z},B_1 b^{r_{B_1}},\dots,B_p b^{r_{B_p}}),
\end{equation}
where the RG eigenvalue $r_{B_i}$ of $B_i$ is given by
\begin{equation}\label{eq:rB}
r_{B_i}\equiv d_{B_i}-\gamma^*_{B_i}
\end{equation}
with $d_{B_i}$ being the na\"ive dimension of $B_i$~\cite{Zhong06,Zhong2}.

To determine $r_{B_i}$, note that the $t$-dependence in Eq.~(\ref{eq:RGeqNew}) results both explicitly from $M$ itself and implicitly from $H$ for the driving. So, one has formally~\cite{Zhong2}
$\partial_{t}=\partial_{t}+\left(\partial_{t}H\right)\partial_{H}$ and $ \partial_{B_i}=\left(\partial_{B_i}H\right)\partial_{H}$.
Substituting them into Eq.~(\ref{eq:RGeqNew}) and comparing the outcome with Eq.~(\ref{eq:RGeq}), we find, at the fixed point,
\begin{equation}\label{eq:expBi}
\varsigma^* \left(\partial_{\ln t}\ln H\right)+\sum_i \gamma_{B_i}^*\left(\partial_{\ln B_i}\ln H\right)=\frac{1}{2}\gamma_{\phi}^*.
\end{equation}
This is a single equation of all $\gamma_{B_i}^*$ for a general $H$. Yet, it must be valid at each instant. This solves all $\gamma_{B_i}^*$ and hence $r_{B_i}^*$ from Eq (\ref{eq:rB}) by comparing similar terms as can be seen from the examples below. Note that all $\gamma_{B_i}^*$ are determined by $\gamma_{\phi}^*$ and $\varsigma^*$ or $\eta$ and $z$ from Eq.~(\ref{eq:sol2exp}). So are all $r_{B_i}^*$. In other words, the usual static and dynamic critical exponents are sufficient for the driven nonequilibrium critical phenomena as has been pointed out.

It will be seen later on that $r_{B_i}^*$ can also be found from direct scale transforms among $H$, $t$, and $B_i$, similar to the linear case~\cite{Zhong06} without explicitly solving Eq.~(\ref{eq:expBi}).

\subsection{\label{tscr}Time scale and crossover}
We now discuss briefly the meaning associated with the parameters of the driving.

From the scaling form~(\ref{eq:Mtau}), one can identify the equilibrium correlation time $t_{\rm eq}\sim |\tau|^{-\nu z}$ and a time scale pertinent to the field $t_H\sim |H|^{-\nu z/{\beta\delta}}$. Both time scales diverge as expected at the exact critical point $\tau=0$ and $H=0$ but are tamed by a finite $\tau$ or $H$.


Similarly, for each parameter, one finds that there is an associated time scale $t_{B_i}$ asymptotically proportional to $B_i^{-z/r_{B_i}}$. In the scaling regime, the shortest long time scale controls the evolution of the system. If $t_{B_i}$ is just such a time scale, the scaling form is then
\begin{equation}\label{eq:MBi}
M=B_i^{\beta/\nu r_{B_i}}f_{B_i}(tB_i^{z/r_{B_i}},B_j B_i^{-r_{B_j}/r_{B_i}},\dots)
\end{equation}
from Eq.~(\ref{mbb}), where $j\neq i$ and $f_{B_i}$ is the associated scaling function. Equation~(\ref{eq:MBi}) implies that physical observables can be rescaled by $B_{i}$ in the critical regime. It is the generalization of the FTS and we shall also refer it as an FTS form. Each argument in $f_{B_i}$ must be vanishingly small to ensure its regularity. This means that $t_{B_i}\ll t_{\rm eq}$ and $t_{B_i}\ll t_{B_j}$ for all $j\neq i$ consistently.

If conditions change such that another time scale, $t_{B_j}$ say, becomes the shortest. In this case, it is now the dominant time scale and governs the leading singularity. Accordingly $f_{B_i}$ is singular near the critical point and behaves as $(B_j B_i^{-r_{B_j}/r_{B_i}})^{\beta/{\nu r_{B_j}}}$ in order to cancel the original singularity. This is a crossover from the regime governed by $t_{B_i}$ to that by $t_{B_j}$. For a general driving with several parameters, such phenomena can be rich.

Moreover, we shall see in the following that there exist time scales that are determined by more than one independent parameter. One case is the first expansion coefficient in $t$ for a general driving. It can be a dominant time scale. By contrast, some time scales may only be transient and never dominate. Although finding the dominant time scale is sometimes not easy, we shall see that the present theory still describes the driven nonequilibrium critical phenomena well.

In addition, there exist crossovers to regimes that are specified by other parameters than $B_i$~\cite{Zhong1,Zhong2}. For example, when $\tau$ is large or $t_{\rm eq}$ dominates, there is a crossover to the adiabatic or (quasi-)equilibrium regime that is governed by it and is described by the scaling form~(\ref{eq:Mtau}) with all $B_i$ present. Similar results can be obtained by other relevant parameters such as $L$. We shall not pursue them further in the following.

\section{\label{sec:initial}Initial Conditions}
In this section, we focus on the effect of initial conditions on driving.
Remember the dynamic equation~(\ref{eq:dye}) is a first-order stochastic differential equation. So, mathematically initial conditions are necessary. An initial condition contains two parts: a starting field $H_{\rm in}$, which characterizes the distance to the critical point, and an initial state distribution $\mathcal {P}_{\rm in}$ of the order parameter $\phi$, or equivalently, all orders of moments of $\mathcal{P}_{\rm in}$. In the previous section, $H_{\rm in}$ is far away from the critical point and thus the initial condition plays no role no matter whether it is in equilibrium or not. By contrast, near the critical point, a nonequilibrium initial condition with correlations \emph{shorter} than the equilibrium ones at $H_{\rm in}$ results in the critical initial slip~\cite{Janssen} even when $H$ is varied linearly~\cite{Huang2}. Here, we consider the effect of initial conditions that have \emph{longer} correlations than the driving ones and that can be in either equilibrium or nonequilibrium.

In general, an initial condition changes with coarse graining. Consequently, upon suppressing other scales, the FTS form including the initial condition is
\begin{equation}\label{eq:FTSIC}
\begin{split}
M=t_D^{-\beta/{\nu z}}f_D(Ht_D^{\beta\delta/{\nu z}},H_{\rm in} t_D^{\beta\delta/{\nu z}},V\left(t_D,\mathcal{P}_{\rm in}\right)),
\end{split}
\end{equation}
where $t_D$ is the dominant time scale of the driving, $H_{\rm in} t_D^{\beta\delta/{\nu z}}$ characterizes the rescaled initial distance to the critical point, and $V\left(t_D,\mathcal{P}_{\rm in}\right)$ is a universal characteristic function describing the rescaled initial distribution, a generation of the critical characteristic function, $U(b,M_{\rm in})$, for an initial state with an arbitrary $M_{\rm in}$ and vanishing correlations~\cite{Zheng,Yin2}, for which $V$ returns to $U$. From Eq.~(\ref{eq:FTSIC}), the FTS regime controlled by $t_D$ satisfies $|H|t_D^{\beta\delta/{\nu z}}\ll1$ and falls within $|H|\ll \hat{H}$, where $\hat{H}\sim t_D^{-\beta\delta/{\nu z}}$ represents the boundary of FTS regime.

If $\left|H_{\rm in}\right|t_D^{\beta\delta/{\nu z}}\gg 1$, $H_{\rm in}$ locates in the adiabatic regime. Accordingly, an initial state either in equilibrium or in nonequilibrium decays exponentially to the equilibrium one quickly with the driving and the initial condition is irrelevant. One can then simply start a driving just beyond the FTS regime with an equilibrium distribution $\mathcal{P}_{\rm eq}(H)$ at $|H|\gtrsim\hat{H}$.

If $\left|H_{\rm in}\right|t_D^{\beta\delta/{\nu z}}\ll 1$ to the contrary, $H_{\rm in}$ lies in the FTS regime. As a result, the information of initial state distribution cannot be ignored. In this case, how the initial condition affects the evolution depends on $\mathcal{P}_{\rm in}$. In the following two subsections, we consider two simple cases that will be used in later sections.

\subsection{Equilibrium initial conditions}
When the initial state is an equilibrium state, the critical initial slip does not matter. In this case, $\mathcal{P}_{\rm in}=\mathcal{P}_{\rm eq}(H_{\rm in})$ and is determined solely by $H_{\rm in}$ as $\tau=0$, e.g., $M_{\rm in}\sim \left|H_{\rm in}\right|^{1/\delta}$ for small $|H_{\rm in}|$. So, $V$ can be expressed by $H_{\rm in}t_D^{\beta\delta/{\nu z}}$. That $H_{\rm in}$ lies in the FTS regime implies $t_D<|H_{\rm in}|^{-{\nu z}/\beta\delta}\sim t_{H_{\rm in}}$. This indicates that the correlation time of the equilibrium initial condition is longer than the driving time $t_D$, and so is the correlation length of the initial condition. In this case, Eq.~(\ref{eq:FTSIC}) becomes
\begin{equation}\label{eq:H0S}
M=\left|H_{\rm in}\right|^{1/\delta}f_{H_{\rm in}}(H/H_{\rm in},t_D |H_{\rm in}|^{{\nu z}/\beta\delta}),
\end{equation}
and the initial condition dominates within $t_{H_{\rm in}}$ even though $t_D< t_{H_{\rm in}}$. The reason is that here the longer scales stemming from the initial condition exist there and thus dominate the short ones that are still setting up. Once the latter has done, the driving $t_D$ takes over.


\subsection{\label{sec:piecewise}Nonequilibrium initial conditions: Continuous piecewise driving}
We next study a specific nonequilibrium initial condition that has longer correlations than the driving scales.

To this end, consider a process with the following two steps. First, start from the adiabatic regime $|H_{\rm in}|>|\hat{H}|$ with a certain form of driving $H_{\rm st_1}$, whose dominant time scale is $t_{D_1}$, and stop at $H_1$ inside the FTS regime of $t_{D_1}$, or $|H_1| t_{D_1}^{\beta\delta/{\nu z}}\ll 1$. This generates a nonequilibrium distribution $\mathcal{P}_1$. Its dominating shortest scale is completely determined by $t_{D_1}$ according to the theory. Second, just at $H_1$, change the form of driving to $H_{\rm st_2}$. This puts $\{H_1,\mathcal{P}_1\}$ at the initial condition of the second step.

If the dominant time scale of the second step $t_{D_2}\ll t_{D_1}$, $H_1$ then satisfies $|H_1| t_{D_2}^{\beta\delta/{\nu z}}\ll|H_1| t_{D_1}^{\beta\delta/{\nu z}}\ll 1$, so that it falls also inside the FTS regime of $t_{D_2}$. This condition eliminates the initial slip of the increase of $M$ too. The scaling form of the second step is thus
\begin{equation}\label{eq:piece2'}
M=t_{D_1}^{-\beta/{\nu z}}f_{\rm st_2}(Ht_{D_1}^{\beta\delta/{\nu z}},
H_1 t_{D_1}^{\beta\delta/{\nu z}},t_{D_2}/t_{D_1}),
\end{equation}
which is dominated by the initial state characterized by $t_{D_1}$ rather than by $t_{D_2}$ from the driving at work, where we have dropped all other possible scales.

\section{\label{sec:examples}Specific forms of driving}
We now apply the results from the last two sections to some specific examples of driving.

\subsection{\label{sec:monomial}Monomial driving}
Consider a driving in a monomial form~\cite{qkz5,protocol,Zhong1,Sandvik}
\begin{equation}\label{eq:HRn}
H(t, R_n)=R_n t^n.
\end{equation}
Without loss of generality, we assume $R_n>0$ in order to simplify the following expressions. Note that the critical point lies exactly at $t=0$ and $H=0$.

Substituting Eq.~(\ref{eq:HRn}) into Eq.~(\ref{eq:expBi}), we have
\begin{equation}\label{gammaRn}
\gamma_{R_n}^*=-n\varsigma^*+\gamma_{\phi}^*/2.
\end{equation}
Using Eq.~(\ref{eq:sol2exp}) and the na\"{\i}ve dimension of $R_n$~\cite{Zhong2},
$d_{R_n}=(d+2)/2+2n$,
which is the difference between the dimensions of $H$ and $t^n$, one finds
\begin{equation}\label{eq:rn}
r_{n}=d_{R_n}-\gamma_{R_n}^*=\beta\delta/\nu+nz.
\end{equation}

We can also reach Eq.~(\ref{eq:rn}) from the scale transforms of $H$ and $t$ similar to the FTS for linearly varying field~\cite{Zhong1,Zhong2}. After a scale transform,
$H'=Hb^{\beta\delta/\nu}$ and $t'=tb^{-z}$
from Eq.~(\ref{eq:Mb}). The definition of $r_{n}$ in Eq.~(\ref{mbb}), viz., $R_n'=R_nb^{r_{n}}$, and Eq.~(\ref{eq:HRn}) then result directly in Eq.~(\ref{eq:rn}), since Eq.~(\ref{eq:HRn}) is also valid when coarse grained, which can also be regarded as a definition of $R_n'$.

From Eq.~(\ref{eq:rn}), for $R_n$ to be relevant, $r_{n}>0$, i.e., $n>-\beta\delta/{\nu z}$. When $n=1$, the FTS for linearly varying field is recovered~\cite{Zhong1,Zhong2}.

Since $R_n$ is the only parameter of $H$, there is only one FTS regime and its FTS form reads
\begin{equation}\label{eq:Rn}
M=R_n^{\beta/{\nu r_{n}}}f_n^t(tR_n^{z/r_{n}})
\end{equation}
directly from Eq.~(\ref{eq:MBi}), or,
\begin{equation}\label{eq:RnH}
M=R_n^{\beta/{\nu r_{n}}}f_n^H(HR_n^{-\beta\delta/{\nu r_{n}}}),
\end{equation}
where we have simplified the subscripts. If we include other parameters such $\tau$ and $L$, we can have crossovers to other regimes. However, as pointed out in Sec.~\ref{tscr}, we shall not consider them further.

There exists a unique singularity stemming from the peculiar property of the scaling functions for a nonlinear driving. Although $tR_n^{z/r_n}=(HR_n^{-\beta\delta/\nu r_n})^{1/n}$, one usually does not care for the exponent and freely applies either Eq.~(\ref{eq:Rn}) or Eq.~(\ref{eq:RnH}) to describe the process, believing that no new singularity will occur except for the confluent ones~\cite{Wegner}. This is not true for $n>1$ however. To see this, note that in general, a scaling function can be Taylor expanded near a critical point. We find, however, that only the expansion of $f_n^t$ can describe $M$ near the critical point, whereas that of $f_n^H$ cannot. This is a manifest of nonequilibrium. It may arise from the fact that the evolution is with the time but not with the field and thus the RG equation for $t$ is better than for $H$ in this case. Yet, substituting $t=(H/R_n)^{1/n}$ into the expansion of the former works well for the latter. This indicates that $f_n^H$ is singular at $H=0$ for $n>1$.

Moreover, the singularity of $f_n^H$ leads to a new leading singularity for the susceptibility $\chi$ at $H=0$, which is
\begin{equation}\label{eq:chi}
\chi=R_n^{-\frac{\gamma}{\nu r_n}}\left[\frac{a_1^{\gtrless}}{n} \left(H R_n^{-\frac{\beta\delta}{{\nu r_{n}}}}\right)^{1/n-1}+\dots\right],
\end{equation}
where $\gamma=\beta(\delta-1)$, $a_1$ is an expansion coefficient independent on both $R_n$ and $H$, and the superscripts $\gtrless$ represent the expansions for $t>0$ and $t<0$, respectively. When $n>1$, $\chi$ diverges at the critical point $H=0$ and $R_n=0$ as $H^{(n-1)/n}$, even in the FTS regime, though it collapses well for different $R_n$ in the plane of $\chi R_n^{\gamma/\nu r_n}$ versus $HR_n^{-\beta\delta/{\nu r_{n}}}$ as Eq.~(\ref{eq:chi}) indicates. Note that in equilibrium, $\chi\sim H^{-\gamma/\beta\delta}$ near $H=0$ but changes to the present nonequilibrium one once $R_n$ is finite, in which case the transition occurs near $HR_n^{-\beta\delta/{\nu r_{n}}}\sim1$ rather than at $H=0$. So, the singularity is all due to the driving as only $n$ is involved. In addition, the leading singularity of $\chi$ also turns into $R_n^{-\beta(\delta-n)/n\nu r_n}$, whose exponent changes sign for $n>\delta$.

The monomial driving with $n\neq 1$ can improve the adiabaticity of a transition~\cite{Barankov}, but requires a better experimental control since it has to be nonlinear exactly at the critical point~\cite{revqkz1}. In the following, we shall see that polynomial forms can be a better approximation.

\subsection{\label{sec:polynomial}Polynomial driving}
Suppose a quadratic driving has small deviations $\delta t$ and $\delta h$ from the zero point, which usually can not be avoided in experiment. Then
\begin{equation}\label{eq:deltat1}
H=R_2(t+\delta t)^2+\delta h.
\end{equation}
For such a driving, we can expand it about the critical point at $H=0$, which lead to a polynomial driving
$H=R_2 t'^2 + R_1 t'$ with $t'=t-t_o$, where $H(t_o)=0$ and $R_1$ is the coefficient of the linear term. Moreover, generally such a driving crosses or approaches $H=0$ several times. In the following, we shall study a driving that crosses $H=0$ only once and multiple times separately.

\subsubsection{\label{sec:polynomial1}Single trans-critical driving}
We first discuss the case in which a driving crosses $H=0$ only once.

The dominant time scale in such a polynomial driving turns out to be quite simple; it is just among all the $t_{R_i}$. So, all we need is to compare and find the smallest of them. To be concrete, consider
\begin{equation}\label{eq:HR1R3}
H(R_1,R_3,t)=R_1 t+R_3 t^3
\end{equation}
with positive $R_1$ and $R_3$ without loss of generality to ensure $t=0$ is the only solution at the critical point $H=0$. From Eq.~(\ref{eq:expBi}), one can see that Eqs.~(\ref{gammaRn}) and (\ref{eq:rn}) for $n=1$ and $n=3$ are the solutions to $\lambda_{R_1}^*$ and $\lambda_{R_3}^*$ and hence $r_1$ and $r_3$, respectively. These results are natural from the direct method of scale transforms as both monomial terms scale with the magnetic field.

As there are two parameters, one has two time scales and two different FTS regimes. Their scaling forms are
\begin{equation}\label{eq:R1R3-1}
M=R_1^{\beta/{\nu r_1}}f_{R_1}(tR_1^{z/r_1},R_3 R_1^{-r_3/r_1}),
\end{equation}
or
\begin{equation}\label{eq:R1R3-2}
M=R_3^{\beta/{\nu r_3}}f_{R_3}(tR_3^{z/r_3},R_1 R_3^{-r_1/r_3})
\end{equation}
from Eq.~(\ref{eq:MBi}). The first form describes the regime when $R_3 R_1^{-r_3/r_1}\ll 1$, or, $t_{R_1}\ll t_{R_3}$, i.e., $t_{R_1}$ dominates in the critical region. By contrast, when $R_3 R_1^{-r_3/r_1}\gg 1$, or, $t_{R_1}\gg t_{R_3}$, the scaling function $f_{R_1} \sim (R_3 R_1^{-r_3/r_1})^{\beta/{\nu r_3}}$ and crosses over to Eq.~(\ref{eq:R1R3-2}) dominated by $t_{R_3}$.

In general, the power $n_i$ of each term in the polynomial driving need not be an integer, as in the monomial case, but it must satisfy $n_i>-\beta\delta/{\nu z}$ to keep $R_{n_i}$ relevant.

\subsubsection{\label{sec:polynomial2}Multi-trans-critical driving}
Here we focus on the case in which the process crosses the critical point $H=0$ several times.

\begin{figure}
  \centering
  \includegraphics[width=\columnwidth]{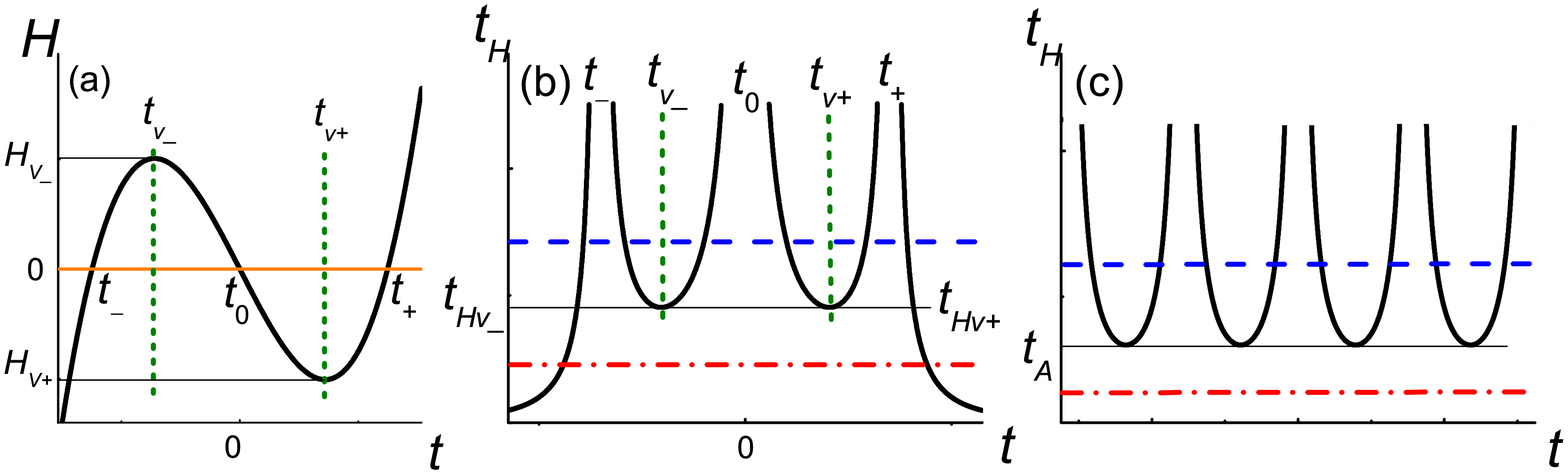}\\
  \caption{(Color Online) (a) Generic $H=R_3 t^3-R_1 t$ curve. (b) Schematic picture of its $t_{H}$ and other time scales versus $t$. (c) Schematic picture of $t_{H}$ and other time scales versus $t$ for the sinusoidal driving $H=A\sin{\Omega t}$. The blue dash lines correspond to the case in which the initial state is dominant, while the red dash-dot lines to the case in which equilibrium can be achieved. See the text for details.}\label{fig:mtth}
\end{figure}
Consider
\begin{equation}\label{eq:mHR1R3}
H=R_3 t^3-R_1 t
\end{equation}
with $R_3>0$ and $R_1>0$ for simplicity. The driving changes direction twice at $t_{v_\pm}=\pm\sqrt{R_1/{3R_3}}$ and crosses the critical point at three instants: $t_0=0$ and $t_\pm=\pm\sqrt{R_1/R_3}$, see Fig.~\ref{fig:mtth}(a). Accordingly, we can divide the process into three parts: 1, $t=(-\infty,t_{v_-}]$, 2, $t=(t_{v_-},t_{v_+}]$, and 3, $t=(t_{v_+},+\infty)$, such that $H$ crosses $H=0$ only once in each part. Figure~\ref{fig:mtth}(b) shows the three divergent peaks and the two valleys of $t_H$. If the dominant driving time scale $t_D\gg t_{H_{v_{\pm}}}$, the time scale at the valleys, the system can equilibrate near $t_{v_\pm}$ and thus the three parts can be treated separately. Conversely, when $t_D\ll t_{H_{v_{\pm}}}$, it stays in the FTS regime even at the valleys and thus the initial conditions are important.

To be specific, we now expand the driving field about each $H=0$. Let us start from $t<0$ outside the FTS regime. The initial condition then plays no role for the first part. Near $t_-$, let $t'=t-t_-$,
\begin{equation}\label{eq:mHt-}
H=R_3 t'^3-3\sqrt{R_1 R_3}t'^2+2R_1 t'.
\end{equation}
A quadratic term $R_2\equiv 3\sqrt{R_1 R_3}$ emerges. As $t_{R_2}/t_{R_1}\sim(R_1 R_3^{-r_1/r_3})^{r_3/{2r_2}}$ and $t_{R_3}/t_{R_2}\sim(R_1 R_3^{-r_1/r_3})^{1/2}$, if $R_1 R_3^{-r_1/r_3}\gg 1$, $t_{R_1}\ll t_{R_2} \ll t_{R_3}$ and $R_1$ dominates; if $R_1 R_3^{-r_1/r_3}\ll 1$, the relation among the three time scales reverses and $R_3$ rules the game. Therefore, no new generated term can dominates except the existing ones.

For $t_0$, no expansion is necessary but we have to consider initial conditions. When $R_1 R_3^{-r_1/r_3}\gg 1$, $R_1$ again dominates.
So, $t_{R_1}/t_{H_{v_{\pm}}}\sim (R_1R_3^{-r_1/r_3})^{\nu z r_3/2\beta\delta r_1}\gg1$ and thus $t_{R_1}\gg t_{H_{v_{\pm}}}$. Therefore, $t_{v_-}$ locates outside the FTS regime and Part 2 can be treated separately. The FTS form in this case is given by Eq.~(\ref{eq:R1R3-1}).
When $R_1 R_3^{-r_1/r_3}\ll 1$ and $R_3$ dominates, in contrast, the initial state is important. From the discussion of $t_-$, the initial state is also dominated by $t_{R_3}$. So the FTS form in this case should be described by a form similar to Eq.~(\ref{eq:piece2'}). However, as $t_{v_-}$ is also determined by $R_1$ and $R_3$, $t_{v_-}R_3^{z/r_3}$ can be reduced to $R_1R_3^{-r_1/r_3}$. Accordingly, the scaling form now resembles Eq.~(\ref{eq:R1R3-2}), with the $R_3$ factor stemming from the initial state from $t_-$. Similar analysis can be applied to the third part $t_+$.

We can also understand roughly the effects of the initial condition from $t_0$, $t_{\pm}$, and $t_{v_{\pm}}$ themselves. For $R_1 R_3^{-r_1/r_3}\gg 1$, they are far apart and so there is sufficient time for the system to equilibrate. As a result, the initial condition is irrelevant. While for $R_1 R_3^{-r_1/r_3}\ll 1$, they are close to each other and the initial condition has to be taken into account.

\subsection{\label{sec:sin}Sinusoidal driving}
Sinusoidal driving is widely used in experimental and theoretical studies, both continuous \cite{sin6,lightscatter} and first-order phase transitions \cite{sin1,sin2,sin3,sin4,sin5,sinf1}. A dynamic phase transition was reported in the kinetic Ising model under a time-dependant oscillating field \cite{sin1}. This may be an example of the resonance interaction. In addition, sinusoidal functions are fundamental in Fourier analysis, so it is instructive to study sinusoidal driving.

Consider
\begin{equation}\label{eq:Hsin}
H(A,\Omega,t)=A\sin{\Omega t},
\end{equation}
where $A$ is the amplitude and $\Omega$ is the angular frequency of the driving. The RG result, Eq.~(\ref{eq:expBi}), indicates that $\lambda_{\Omega}^*=-\varsigma^*$ and $\lambda_{A}^*=\gamma^*/2$ and thus $r_{\Omega}$ and $r_A$ are simply $z$ and $\beta\delta/\nu$, respectively. This is obvious because $A$ must transform as $H$ and $\Omega$ as $t^{-1}$ from Eq.~(\ref{eq:Hsin}).

Accordingly, the scaling forms for $\Omega$ and $A$ dominating are respectively
\begin{eqnarray}
M(t,A,\Omega)&=&\Omega^{\beta/{\nu z}}f_\Omega\left(\Omega t,A\Omega^{-\beta\delta/{\nu z}}\right),\label{eq:sinw}\\
M(H,A,\Omega)&=&A^{1/\delta}f_A\left(HA^{-1},\Omega A^{-\nu z/{\beta\delta}}\right),\label{eq:sinA}
\end{eqnarray}
since $t_{\Omega}\ll t_{A}$, i.e., $\Omega^{-1}\ll A^{-\nu z/\beta\delta}$ or $A\Omega^{-\beta\delta/{\nu z}}\ll1$ for the former and vice versa for the latter. We can of course choose $H$ and $t$ as their respective parameters.

But there exists yet another way to study the driving. We can expand the driving near the critical point at $t=0$ and $H=0$ and utilize the theory for the polynomial in Sec.~\ref{sec:polynomial}. When the first-order term dominates, we find the scaling form
\begin{equation}\label{eq:sinR1}
\begin{split}
M=(A\Omega)^{\beta/{\nu r_1}}f_{A\Omega}\left(H(A\Omega)^{-\beta\delta/{\nu r_1}},\Omega A^{-{\nu z}/\beta\delta}\right),
\end{split}
\end{equation}
because all the rescaled higher-order terms are simply reduced to $A^{-1}\Omega^{\beta\delta/{\nu z}}$. To be consistent, $t_{A\Omega}\sim (A\Omega)^{-z/r_1}\ll t_H$ and $\Omega A^{-\nu z/\beta\delta}\ll1$. One sees therefore that there is a new dominating timescale $t_{A\Omega}$ determined by two parameters, whose combination cannot be identified directly from the field itself!

In order to reveal the relationship among Eqs.~(\ref{eq:sinw}) to (\ref{eq:sinR1}), we notice that
\begin{equation}
\frac{t_{A\Omega}}{t_A}\sim \left(A\Omega^{-\frac{\beta\delta}{\nu z}}\right)^{\frac{\nu z^2}{\beta\delta r_1}},\quad
\frac{t_{A\Omega}}{t_\Omega}\sim  \left(\Omega A^{-\frac{{\nu z}}{\beta\delta}}\right)^{\frac{\beta\delta}{\nu z}}\label{tawta}.
\end{equation}
Therefore, for $\Omega A^{-\nu z/\beta\delta}\ll1$, $t_A\ll t_{A\Omega}\ll t_\Omega$, while for $\Omega A^{-\nu z/\beta\delta}\gg1$, $t_\Omega\ll t_{A\Omega}\ll t_A$. This appears to indicate that $t_A$ would dominate the former regime and $t_\Omega$ the latter. In other words, no regime dominated by $t_{A\Omega}$ would emerge.
In fact, it is $t_{A\Omega}$ that dominates the former regime and the initial state that governs the latter. $t_A$ is only transient and never dominant.

To see this, note that $t_A$ is just the minima of $t_H$ from Fig.~\ref{fig:mtth} (c). Accordingly, it is only a transient time scale in the sense that it only appears at the instant when $H$ assumes its maxima or minima and thus cannot be a constantly dominating time scale. Therefore, for $\Omega A^{-\nu z/\beta\delta}\ll1$, although $t_{A\Omega}\gg t_A$ from Eq.~(\ref{tawta}), corresponding to the blue dashed line in Fig.~\ref{fig:mtth} (c), $t_{A\Omega}$ is the dominant scale. In addition, the system can equilibrate at the valleys and the hysteresis loops are saturate. If $\Omega A^{-\nu z/\beta\delta}\gg1$, although $t_{\Omega}$ is the shortest, the system always stays in the FTS regime and the initial condition dominates; only after the initial state decays away can $t_{\Omega}$ take over and dominate. Moreover, the time scales of the higher expansion coefficients are shorter than that of the linear term and the expansion method is invalid.

As the initial state dominates for $\Omega A^{-\nu z/\beta\delta}\gg1$, the hysteresis loops become unsaturate.
If we start the process in equilibrium at $H_{\rm in}=-A$ and $\mathcal{P}_{\rm in}=\mathcal{P}_{\rm eq}(-A)$, the scaling in this regime is described by
\begin{equation}\label{eq:sinH0}
M= A^{1/\delta}f_{H_{\rm in}=-A}(HA^{-1},\Omega^{-1} A^{{\nu z}/\beta\delta}),
\end{equation}
where the subscript differentiates it from Eq.~(\ref{eq:sinA}), as Eq.~(\ref{eq:sinH0}) is a special case of Eq.~(\ref{eq:H0S}) for a specific initial condition used. It is valid for a small $A$ and a large $\Omega$, in opposite to Eq.~(\ref{eq:sinA}), which never dominates.

\subsection{\label{gauss}Gaussian approximation}
As an appreciation of the various timescales associated with a driving and a verification of the above results, in this section, we consider the Gaussian approximation of the model~(\ref{eq:F}) and (\ref{eq:dye}).

In this approximation, the model simplifies to
\begin{equation}\label{eq:gause}
\frac{\partial \langle \phi(k,t)\rangle}{\partial t}=-\lambda[(\tau+k^2) \langle\phi(k,t)\rangle-H(t)\delta({\mathbf k})]
\end{equation}
in the wavenumber ${\mathbf k}$ space, where $\delta$ here is the Dirac delta function. The solution is
\begin{equation}\label{eq:Ggen}
\begin{split}
M(t)=e^{- \tau\lambda (t-t_{\rm in})}M_{\rm in}-g(t_{\rm in})e^{- \tau\lambda (t-t_{\rm in})}+g(t),
\end{split}
\end{equation}
where $M_{\rm in}$ is the initial uniform magnetization. The first two terms in Eq.~(\ref{eq:Ggen}) are the contributions of initial state. They decay exponentially. The third term is the effect of driving and depends on its detail. If $t-t_{\rm in}\sim 0$, the last two terms nearly cancel out and the first term dominates. This means the time is too short for the effect of driving to be significant and the result mainly reflects the property of $M_{\rm in}$. By contrast, when $t-t_{\rm in}\gg 0$, only the driving term survives.

For the polynomial form of driving~(\ref{eq:HR1R3}) but with all $t$ replaced by $\lambda t$ in order to yield a correct time unit, the solution is $
g(\lambda t)=R_3 (\lambda t)^3/\tau-3R_3(\lambda t)^2/\tau^2
+\left(R_1 /\tau+6R_3/\tau^3\right) \lambda t-\left(R_1 /\tau^2+6R_3/\tau^4\right)$.
The equilibrium correlation time $ t_{\rm eq}=(\lambda\tau)^{-1}$ for the parameters chosen, as the critical exponents of the Gaussian model are $\nu=1/2$ and $z=2$. In addition, $r_1=(d+6)/2$ and $r_3=(d+14)/2$ from Eq.~(\ref{eq:rn}) for $\beta=(d-2)/4$ and $\delta=(d+2)/(d-2)$, and hence $\lambda t_{R_1}= R^{-4/(d+6)}$ and $\lambda t_{R_3}=R^{-4/(d+14)}$. These then turn~(\ref{eq:Ggen}) into
\begin{equation}\label{eq:GR1R3-2}
M=t_{R_3}^{-(d-2)/4}f_{G3}\left(t_{\rm eq}/t_{R_3},t/t_{R_3},t_{R_1}/t_{R_3}\right),
\end{equation}
where $f_{G3}(X,Y,Z)=XY^3-3X^2Y^2+(Z^{-2}X+6X^3)Y-(Z^{-2}X^2+6X^4)$.
Equation~(\ref{eq:GR1R3-2}) is a form of Eq.~(\ref{eq:MBi}) and verifies it. We can of course rescale all time scales by $t_{R_1}$.

In the case of sinusoidal driving, the solution is
\begin{equation}\label{eq:Gsin}
g(\lambda t)=\frac{A\sin(\Omega \lambda t-\theta)}{\sqrt{\Omega^2+\tau^2}}
=\frac{At_{\rm eq}\sin(t/t_{\Omega}-\theta)}{\sqrt{1+t_{\rm eq}^2/t_{\Omega}^2}},
\end{equation}
where $\theta=\arctan(\Omega/\tau)$. When $\Omega/\tau=t_{\rm eq}/t_{\Omega}\ll 1$, the period is significantly longer than the correlation time. So, $t_\Omega$ is not relevant and $M$ will saturate at some time in the process. In the region $t\sim 0$, Eq.~(\ref{eq:Gsin}) can be approximated to be $M\sim (A\Omega)^{(d-2)/(d+6)}(t_{\rm eq}/t_{A\Omega})(t/t_{A\Omega})$, consistent with Eq.~(\ref{eq:sinR1}) in which $t_{A\Omega}$ dominates.

When $\Omega/\tau\gg 1$, the situation reverses and the system can not equilibrate during the whole process, which corresponds to the unsaturated case. In this case, $\theta\sim \pi/2$, Eq.~(\ref{eq:Gsin}) approximates to $M\sim A/\Omega=A^{1/\delta}(\Omega^{-1}A^{\nu z/\beta\delta})$ in agreement with Eq.~(\ref{eq:sinH0}) near $t\sim 0$. One sees that no regime is controlled by $t_A$, consistent with the theory.

\section{\label{sec:wArgue}Measurement}
We discuss a possible implication of our results to experiments here.

Experimentally, one often applies a weak driving field to study the property of a system \cite{conductivity,lightscatter}. For example, in~\cite{conductivity}, the authors measured the linear conductivity between $3$Hz to $3$MHz, and found a dynamic scaling near the vortex--glass transition. In~\cite{lightscatter}, the correlations of order parameters are related to light scattering intensities. In both cases the amplitude of the driving is ignored in the scaling function by assuming that the amplitude is small. In theories, the linear response~\cite{Mazenko} for example, one also applies a weak external field to compute the response of a system. But the field is sent to zero after the computation. In experiments, however, the field is always there no matter how small it is. This usually incurs only a small perturbation. But near the critical point where correlations are long ranged, it may be problematic. Recall that in Eqs.~(\ref{eq:sinw}) to (\ref{eq:sinR1}), there exists an additional term containing both $A$ and $\Omega$. Upon omitting this term, the method to obtain critical properties from data collapses, as was done in~\cite{conductivity}, is thus presumably flawed. A na\"{\i}ve way to overcome this from the theory is to vary $A$ with $\Omega$ in a such way that $\Omega A^{-{\nu z}/\beta\delta}$ is fixed.

The problem can also be seen more fundamentally from the fluctuation-dissipation theorem (FDT)~\cite{Mazenko}, which is
\begin{equation}\label{eq:FDT}
\chi\equiv(\partial M/\partial H )=C/T\equiv L^{d}\left\langle (\phi -\langle \phi \rangle)^2\right\rangle/T
\end{equation}
here. It enables one to measure the equilibrium correlation $C$ via the response of a system to its external probes, the susceptibility $\chi$. We shall see unambiguously nonequilibrium behavior and violation of FDT for a small driving near a critical point. Therefore, one cannot obtain accurate correlation functions by measuring the responses and vice versa near the critical point even for a vanishingly small $A$! Nonequilibrium still, we shall find that the scaling law holds between the critical exponents of $\chi$ and $C$. Accordingly, one can still employ $C$ or $\chi$ to estimate the critical exponents as in equilibrium with due attention to the effect of the amplitude.

\section{\label{sec:model}Model and Method}
To verify our results, we study the classical $d=2$ Ising model with nearest-neighbor interaction
\begin{equation}\label{eq:ising}
\mathcal{H}=-J\sum_{\langle i,j\rangle}{s_i s_j}-H\sum_i{s_i},
\end{equation}
where $J>0$ is a coupling constant and $s_i=\pm 1$ is the spin at site $i$. Periodic boundary conditions are applied throughout. The order parameter is defined as
$M=\langle\sum_i{s_i}\rangle/N$ for the $N$ spins as usual.
The critical temperature and critical exponents are known exactly: $T_c=2J/{\log(1+\sqrt{2})}$, $\beta=1/8$, $\delta=15$, $\nu=1$, while the dynamic exponent is chosen as $z=2.1667$~\cite{z}.

We use MC with a single site Metropolis algorithm~\cite{MC}. To minimize the finite size effect, the minimum lattice size chosen is $512\times 512$. The sample sizes are between $500$ to $3000$, resulting in small relative errors to be seen in the error bars displayed. To reduce variables in the scaling functions, all simulations are performed at $T=T_c$, and $M_0\equiv |M(H=0)|$ is frequently used.

\section{\label{sec:result}numerical results}
\subsection{\label{sec:resH0}Initial Condition}
\begin{figure}
\centering
\includegraphics[width=\columnwidth]{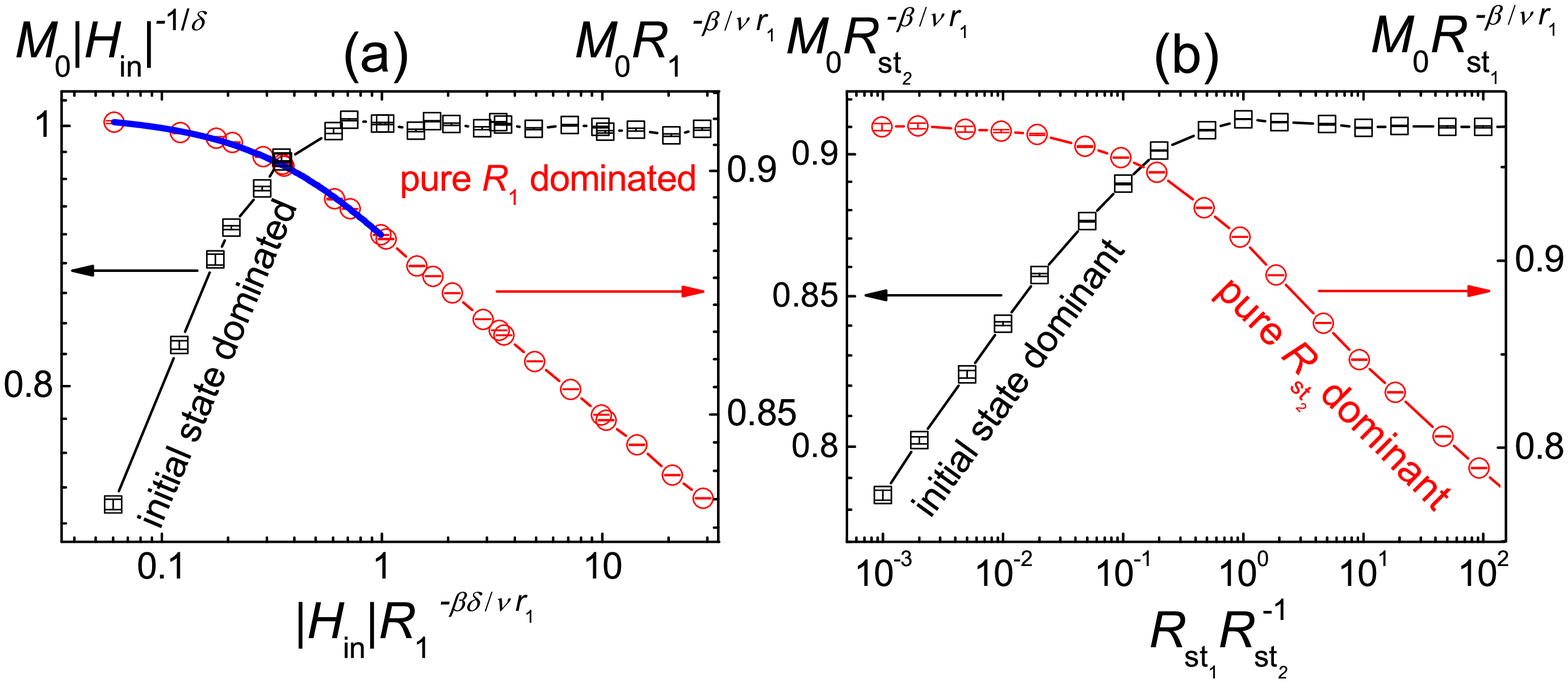}
\caption{(Color online) Effects of (a) equilibrium initial condition and (b) nonequilibrium initial condition of the stepwise linear driving with $t_1 R_{\rm st_1}^{z/r_1}=0.5$. On the right hand side, the black curves are flat showing the irrelevance of the initial state; while the red curves have slopes of $-0.0674(3)$ in (a) and $-0.0294(3)$ in (b), consistent with the theoretic value $-1/\delta=-0.0666$ and $-\beta/{\nu r_1} = -0.0309$, respectively. On the left hand side, the black curves depend on the initial conditions and the red curves tend to be flat, showing clearly the existence of the initial state dominated regime. Correspondingly, the slopes of the black curves become $0.0538(1)$ for the leftmost three data in (a) and $0.0311(2)$ in (b), close to the theoretical values $1/\delta$ and $\beta/{\nu r_1}$, respectively. The thick blue curve in (a) is a fit to the expansion of the scaling function in Eq.~(\ref{eq:H0S}) to order 2. Thin lines connecting symbols are only a guide to the eye.}\label{fig:R1R2}
\end{figure}
Figure~\ref{fig:R1R2} (a) shows $M_0$ of a linear driving starting from an equilibrium initial condition at different $|H_{\rm in}| R_1^{-\beta\delta/{\nu z}}$ for $H_{\rm in}<0$. One sees clearly a crossover from an $R_1$ dominated regime described by Eq.~(\ref{eq:FTSIC}) in the absence of $V$ to an initial-state dominated regime described by Eq.~(\ref{eq:H0S}) with $t_D=t_{R_1}$. The good fit both confirms the scaling and demonstrates the regularity of the scaling function. These results show convincingly the effects of the initial condition and the validity of the theory.

\begin{figure}
  \centering
  \includegraphics[width=0.9\columnwidth]{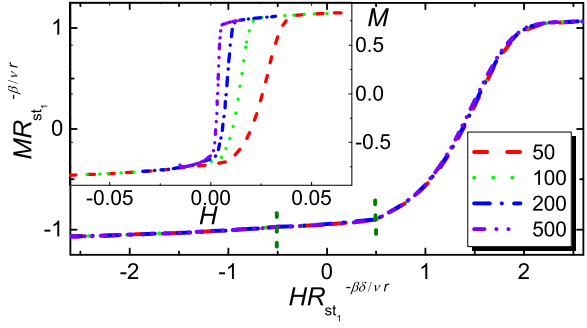}
  \caption{(Color online) Magnetization of the three-step piecewise linear driving. $t_1 R_{\rm st_1}^{z/r_1}=0.5$ and $R_{\rm st_1}/R_{\rm st_2}=0.2$. These choices yield $H_1 R_{\rm st_2}^{-\beta \delta/{\nu r_1}}=0.2370<1$ and so $H_1$ lies inside the FTS regime of $R_{\rm st_2}$. The numbers list the choice of $t_1$. The two vertical dashed lines demarcate the three steps. Inset: Original curves before rescaled.}\label{fig:step}
\end{figure}
Figure~\ref{fig:step} shows the results of the piecewise driving for the nonequilibrium initial condition. A three-step linear driving is simulated. It starts at $H_{\rm in}<-\hat{H}$ of the first driving. For simplicity, we choose $R_{\rm st_3}=R_{\rm st_1}$ and stop the second step at $H_2=-H_1>0$. We also set $H_1=-R_{\rm st_1}t_1$. As a result, the four free parameters of the initial conditions, $t_1$, $t_2$, $R_{\rm st_1}/R_{\rm st_2}$, and $R_{\rm st_2}/R_{\rm st_3}$, are reduced to $t_1$ and $R_{\rm st_1}/R_{\rm st_2}$. In Fig.~\ref{fig:step}, $H_1$ is chosen to fall inside the FTS regime of the second driving, but then $H_2$ lies outside that of the last driving. Also $t_{R_{\rm st_1}}>t_{R_{\rm st_2}}$. Accordingly, the second stage is described by Eq.~(\ref{eq:H0S}), while the other two by Eq.~(\ref{eq:RnH}), though all three stages are rescaled by $R_{\rm st_1}$. The good collapses show the applicability of FTS to this case well.

The nonequilibrium initial conditions dominated regime and its crossover for the driving are displayed in Fig.~\ref{fig:R1R2} (b). Here, we vary the value of $R_{\rm st_1}/R_{\rm st_2}$, which also changes $H_1 R_{\rm st_2}^{-\beta\delta/{\nu r_1}}$. It is clear that the initial state is dominant for $R_{\rm st_1}/R_{\rm st_2}\ll1$, but irrelevant to the opposite. A crossover appears near $R_{\rm st_1}/R_{\rm st_2}\sim 1$. The results show remarkably that, first, the single driving scale does determine all correlations of the system, and second, in the nonequilibrium initial state dominated regime, all correlations still evolves in a concerted way as if the previous driving were still in effect.

\subsection{\label{sec:resMo}Monomial}
\begin{figure}
\centering
\includegraphics[width=\columnwidth]{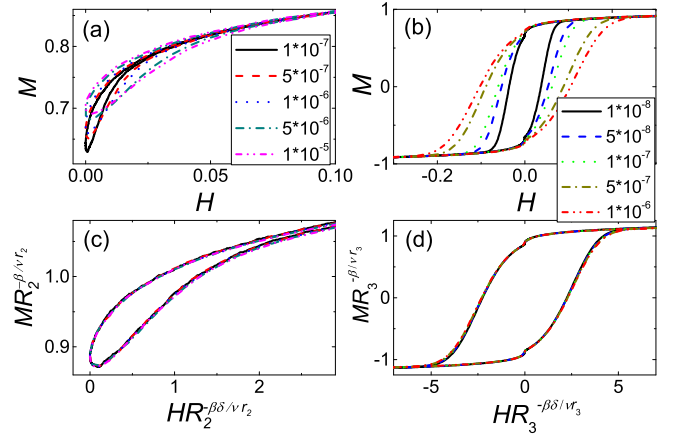}
\caption{(Color online) Hysteresis loops for $H=R_2 t^2$ (left column) and $H=R_3 t^3$ (right column) for various $R_2$ and $R_3$ listed. The original loops (upper panels) collapse well after rescaled (lower panels), verifying Eq.~(\ref{eq:RnH}).}\label{fig:R3}
\end{figure}
In Fig.~\ref{fig:R3}, we verify Eq.~(\ref{eq:RnH}) for $n=2$ and $3$. Although the forms of the driving are qualitatively different as $n=2$ is even but $n=3$ is odd, they both obey the theory.

\begin{figure}[b]
  \centering
  \includegraphics[width=\columnwidth]{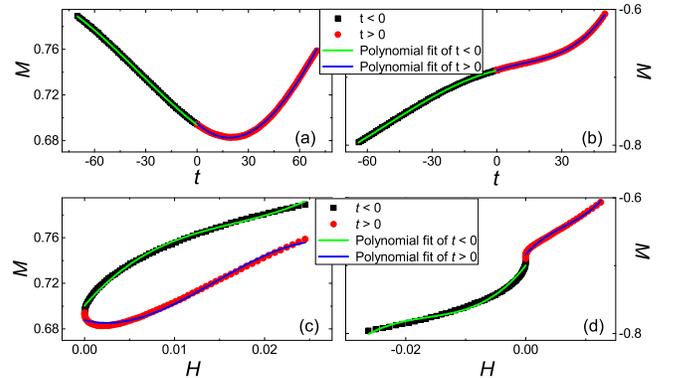}
  \caption{(Color online) $M$ vs $t$ (upper panels) and $M$ vs $H$ (lower panels) for $H=R_n t^n$ with $n=2$ (left panels) and $n=3$ (right panels), respectively. The data are chosen to satisfy $|t|R_n^{z/r_n}<1$ near the critical point and are fitted to polynomials up to order $3$ for $t>0$ and $t<0$, respectively. The goodness of the fits confirms the regular expansion of $f_n^t$ in Eq.~(\ref{eq:Rn}). For (c) and (d), the fits to polynomials up to order $9$ and far smaller ranges are also poor and thus invalid the regular expansion of $f_n^H$ in Eq.~(\ref{eq:RnH}).}\label{fig:thm}
\end{figure}
In Fig.~\ref{fig:thm} we investigate the behavior of $f_n$ in Eqs.~(\ref{eq:Rn}) and (\ref{eq:RnH}) near $H=0$ for $n=2$ and $3$. One sees that the polynomial fits are very good in (a) and (b) but deviate significantly from the data in (c) and (d). This indicates that $f_n$ behaves regularly with respect to $t$ but not to $H$ near the critical point.

\begin{figure}
  \centering
  \includegraphics[width=\columnwidth]{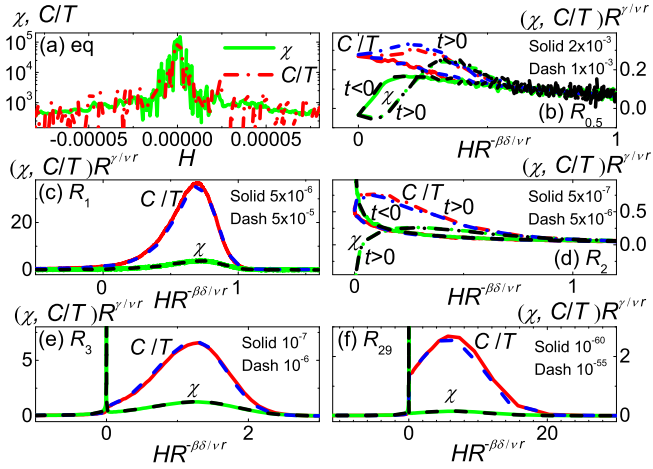}
  \caption{(Color online) (a) $\chi$ and $C/T$ vs $H$ at equilibrium and (b) to (f) their rescaled for the driving $H=R_n t^n$ with (b) $n=0.5$, (c) $n=1$, (d) $n=2$, (e) $n=3$, and (f) $n=29$, respectively. Note that subscripts are absent in $R$ and $r$ of the axis titles. $C/T$ and $\chi$ mark their rescaled curves in red and blue and in green and black, respectively. The solid and dashed curves represent the different rates listed. In (b) and (d), they correspond to $t<0$, while the dash-dotted curves to $t>0$. Note that $\chi$ is negative for $t$ close to $0^+$. As predicted by Eq.~(\ref{eq:chi}), $\chi$ vanishes, is finite, and diverges at $H=0$ for $n<1$ (b), $n=1$ (c), and $n>1$ (d) to (f), respectively. The collapses of $C/T$ are not as good as $\chi$ due to large fluctuations. The two curves in (a) are averaged over 10,000,000 MC steps.}\label{fig:chihT}
\end{figure}
Figure \ref{fig:chihT} shows the rescaled curves of $\chi$ and $C/T$ for two different rates. Note that $\chi$ is negative in (b) and (d) for $t$ close to $0^+$. In equilibrium, however, $\chi$ is related via the FDT~(\ref{eq:FDT}) to $C$, which is nonnegative. So the negative $\chi$ implies definitely that the system is out of equilibrium. It can also be seen that $\chi$ and $C/T$ separate significantly near the peaks for all forms of driving shown. This violation of the FDT is again manifestly a nonequilibrium effect, which maximizes near the peaks where the transition takes place. Yet, the good collapses of both $\chi$ and $C$ indicate that the scaling law between the critical exponents holds even there. One might regard the scaling with the driving field in the absence of a new leading exponent as a kind of adiabaticity in which one replaces the field directly by its time dependent form and the system would just evolve according to it. The present results demonstrate that this is not all.

\subsection{\label{sec:resPo}Polynomial}
\begin{figure}[b]
\centering
\includegraphics[width=\columnwidth]{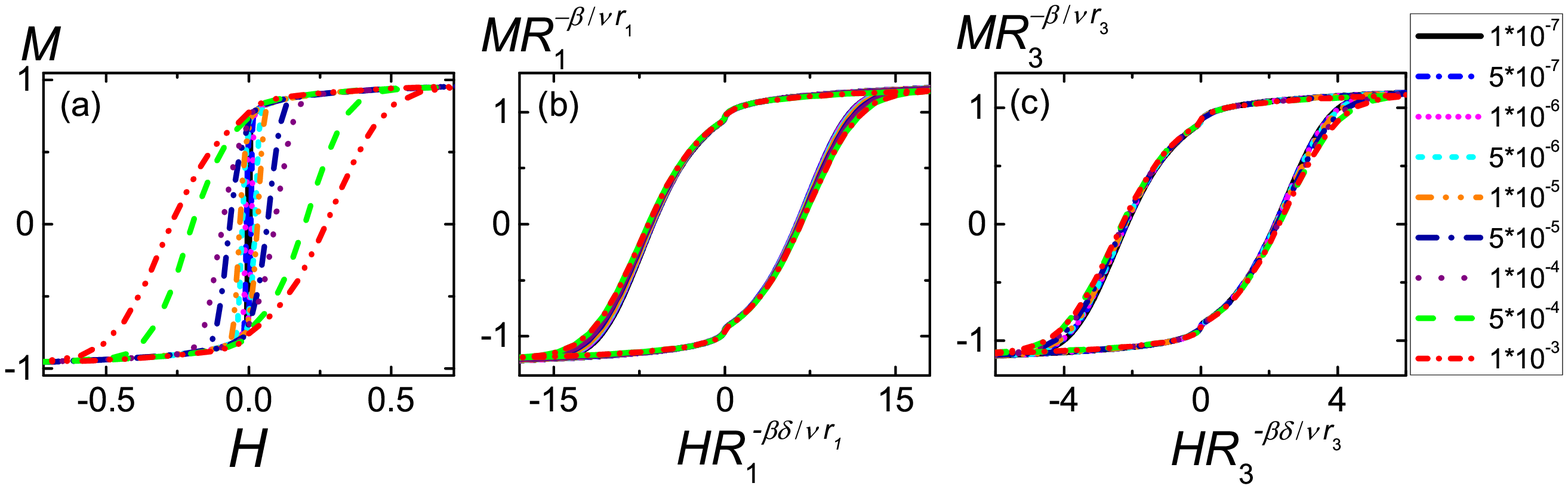}
\caption{(Color online) Hysteresis loops for $H=R_1 t+ R_3 t^3$ with $R_1 R_3^{-r_1/r_3}=0.1$. The legend lists $R_1$ used. (a) is original data, (b) is rescaled by $R_1$, and (c) is rescaled by $R_3$. The collapses get somehow poor when the absolute values of horizontal axis are large, possibly due to corrections to scaling.}\label{fig:R1R3r}
\end{figure}
Figure \ref{fig:R1R3r} depicts the hysteresis loops of the polynomial driving Eq.~(\ref{eq:HR1R3}). According to Eqs.~(\ref{eq:R1R3-1}) and (\ref{eq:R1R3-2}), for fixed $R_1 R_3^{-r_1/r_3}$, all rescaled curves collapsed well onto each other as shown, demonstrating that both $R_1$ and $R_3$ can describe the scaling well.

\begin{figure}
  \centering
  \includegraphics[width=\columnwidth]{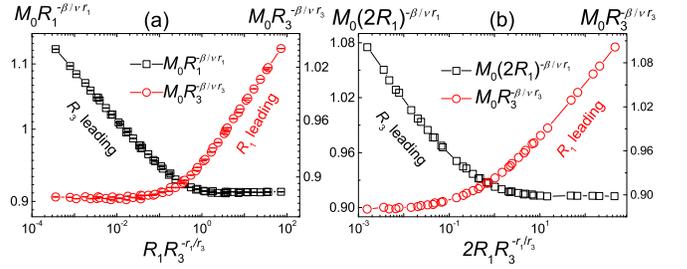}
  \caption{(Color online) Different FTS regimes and their crossover for (a) $H=R_1 t+ R_3 t^3$ and (b) $H=R_3 t^3- R_1 t$ near $t_-$. Black and red curves are the same data rescaled by $R_1$ ($2R_1$ in (b)) and $R_3$, respectively. Each rescaled curve consists of a leading $R_1$ section (where $R_1 R_3^{-r_1/r_3}\ll 1$), a leading $R_3$ section (where $R_1 R_3^{-r_1/r_3}\gg 1$) with different slopes and a crossover between them (where $R_1 R_3^{-r_1/r_3}\sim 1$). The slope of the black (red) curve in the $R_3$ ($R_1$) regime is $-0.0305(1)$ ($0.0309(1)$) in (a) and $-0.0309(1)$ ($0.0299(4)$) in (b), consistent with theoretical absolute value of $\beta/{\nu r_1}$. No error bars appear in (b) as $M_0$ is obtained by interpolating the averaged magnetization curves at the first $H=0$. But they cannot be appreciably larger than those displayed in Fig.~\ref{fig:R1R3m} below. Lines connecting symbols are only a guide to the eye.}\label{fig:R1R3}
\end{figure}
\begin{figure}
  \centering
  \includegraphics[width=\columnwidth]{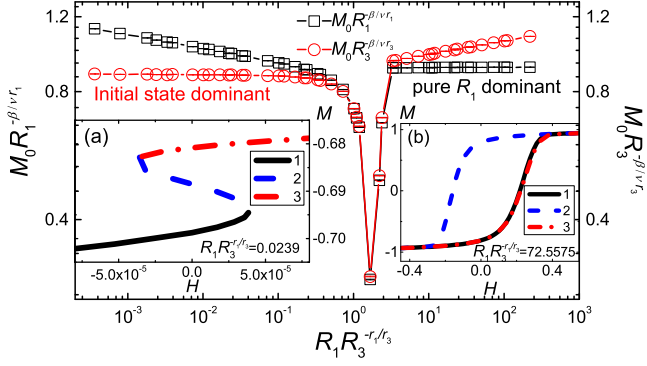}
  \caption{(Color online) Different regimes and their crossover for $H=R_3 t^3 -R_1 t$ near $t_{0}$. Black and red curves are the same data rescaled by $R_1$ and $R_3$, respectively. When $R_1 R_3^{-r_1/r_3}\ll 1$, the red curve is flat, while the black curve has a slope $-0.0320(2)$, consistent with the theoretical value of $-\beta/{\nu r_1}$ from Eqs.~(\ref{eq:R1R3-1}) and (\ref{eq:R1R3-2}). When $R_1 R_3^{-r_1/r_3}\gg 1$, the process can be treated as a pure $R_1$ driving. As a result, the black curve becomes flat, whereas the slope of the red curve is $0.0310(5)$, in agreement with the theoretical value of $\beta/{\nu r_1}$. Lines connecting symbols are only a guide to the eye. Insets: Generic $M$ vs $H$ curves in the two corresponding regimes. Black solid, blue dashed, and red dash-dotted curves represent the 1, 2, and 3 parts of the process, respectively. In (a), the blue curve begins at $H>0$ and $M<0$, which is evidently nonequilibrium. In (b), the curves of the first (black) and the third (red) parts nearly coincide.}\label{fig:R1R3m}
\end{figure}
To investigate the different regimes and their crossover, we again pick the data at $H=0$, so that the scaling function in Eqs.~(\ref{eq:R1R3-1}) and (\ref{eq:R1R3-2}) are only affected by $R_1 R_3^{-r_1/r_3}$. The results of the rescaling are shown in Fig.~\ref{fig:R1R3} (a). Each curve becomes flat when rescaled by the right rescaling variable at its corresponding regime, implying that the effect of the other variable can be ignored in that regime. The good agreement with the theory confirms the latter. A similar result appears for the driving~(\ref{eq:mHR1R3}) near $t_-$, Fig.~\ref{fig:R1R3}(b). This indicates that the new emerged second order term never dominates, though it gives rise to a larger region of crossover by comparing Fig.~\ref{fig:R1R3} (a) and (b). Near $t_0=0$, the graph now appears somehow different as shown in Fig.~\ref{fig:R1R3m}, whose insets demonstrate manifestly the dramatic effect of nonequilibrium initial conditions. For $R_1 R_3^{-r_1/r_3}\ll 1$, the system does not have enough time to relax to equilibrium near $t_{v_-}$. Consequently, the early part affects the later one in contrast to the opposite regime in which all three curves start and end in equilibrium and can thus be treated separately. $M_0$ is thus still negative as seen in Inset (a) and results in the dig as absolute values are used. This indicates that the initial state controls the evolution in this regime. Because the dominant time scale of the first part is $t_{R_3}$, the initial state is again dominated by $R_3$. This is why the two original parameters can describe the scaling well using Eqs.~(\ref{eq:R1R3-1}) and (\ref{eq:R1R3-2}). Nonetheless, separating the process and considering the initial conditions reveal far rich phenomena and physics.

\subsection{\label{sec:resSin}Sinusoidal}
\begin{figure}
\centering
\includegraphics[width=\columnwidth]{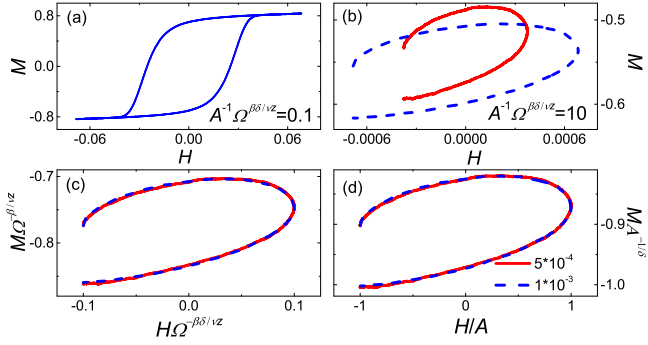}[b]
\caption{(Color online) Hysteresis loops of the sinusoidal driving $H=A\sin{\Omega t}$. (a) For $A^{-1}\Omega^{\beta\delta/{\nu z}}=0.1<1$, $M$ saturates at large $H$. (b) For $A^{-1}\Omega^{\beta\delta/{\nu z}}=10>1$, $M$ does not saturate. (c) and (d) are the rescaled results of (b). (c) is rescaled by $A$ and (d) is rescaled by $\Omega$. The legend lists $\Omega$ used in (b), (c), and (d).}\label{fig:sinloop}
\end{figure}
Here, all simulations are performed with the initial condition $H_{\rm in}=-A$ and $\mathcal{P}_{\rm in}=\mathcal{P}_{\rm eq}(-A)$. We first show the hysteresis loops for the saturated and unsaturated cases in Fig.~\ref{fig:sinloop}. Note that in the unsaturated case, the loops are not close and the range of $M$ is relatively small. This is because equilibrium can not be achieved during the whole process and the initial condition is important. One sees that even though they are not dominated by $A$ and $\Omega$, the hysteresis loops can be rescaled well by them according to Eqs.~(\ref{eq:sinA}) and (\ref{eq:sinw}), respectively.

\begin{figure}
  \centering
  \includegraphics[width=\columnwidth]{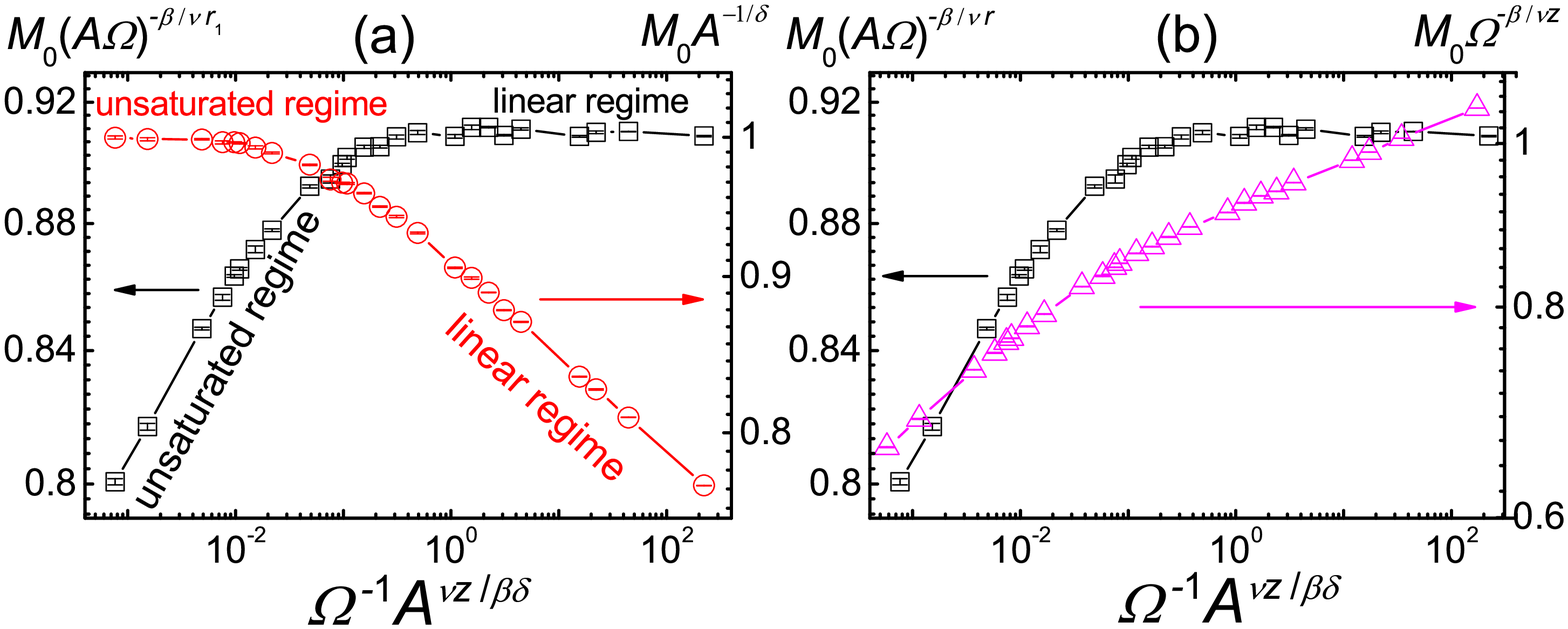}
  \caption{(Color online) Different regimes and their crossover for $H=A\sin{\Omega t}$. Rescaled of the same data (a) by $A\Omega$ (black squares) and $A$ (red circles), and (b) by $A\Omega$ (black squares) and $\Omega$ (magenta triangles). The black curves are identical in (a) and (b). The magenta curve is not flat in any regime, implying that $\Omega$ does not dominate. The black curves become flat at $\Omega^{-1}A^{\nu z/{\beta\delta}}\gg 1$, indicating the dominance of $t_{A\Omega}$ in this regime. Correspondingly, the slopes of red and magenta curves are $-0.0312(2)$ and $0.0559(4)$, in agreement with the theoretic values of $-\beta/{\nu r_1}$ and $\beta/{\nu z}=0.0577$ according to Eqs.~(\ref{eq:sinR1}) and (\ref{eq:sinH0}) and Eqs.~(\ref{eq:sinw}) and~(\ref{eq:sinH0}), respectively. For $\Omega^{-1}A^{\nu z/{\beta\delta}}\ll 1$, the red curve becomes flat and the slopes of the black and magenta curves are $0.0289(5)$ and $0.0269(2)$, consistent with the theoretic values of $\beta/{\nu r_1}$ and $(1/z-1/r_1)\beta/\nu=0.0268$, respectively, showing the importance of the initial condition. Lines connecting symbols are only a guide to the eye.}\label{fig:wmu}
\end{figure}
Figure~\ref{fig:wmu} is the verification of the scaling forms~(\ref{eq:sinR1}), (\ref{eq:sinw}) and~(\ref{eq:sinH0}). One sees that there exists no regime in which $t_{\Omega}$ dominates, even in the regime $\Omega^{-1}A^{\nu z/{\beta\delta}}\ll 1$ in which it is the smallest time scale. In this regime, the initial condition is dominant and the hysteresis loops are unsaturate as Fig.~\ref{fig:sinloop} (b) shows. Accordingly, the leading scaling with $A$ here stems from the initial condition rather than from $t_A$. After the initial condition decays away, $t_{\Omega}$ may take over. In the opposite regime in which $t_A$ is smallest from Eq.~(\ref{tawta}), it is $t_{A\Omega}$ instead of $t_A$ that dominates, as seen in Fig.~\ref{fig:wmu}. In other words, $t_A$ never dominates, as it is a transient scale.  All these confirm well our theory of the sinusoidal driving.

\section{\label{sec:conclusion}Conclusion}
We have studied a generic class of nonequilibrium systems representing by a critical system that possesses a long relaxation time and is weakly driven within a finite time in a form that does not cause resonances but otherwise is arbitrary. An RG theory has been developed to account for such driven nonequilibrium critical phenomena. From the theory, the driving generates finite time scales that can well be shorter than the equilibrium correlation time and thus driving the system far off equilibrium. This creates topological defects of the KZ mechanism. Moreover, the finite time scales can control different regimes and thus crossovers among them can take place when conditions change. These nonequilibrium phenomena are well described in the theory with just the usual static and dynamic critical exponents. Yet, this does not mean that a kind of adiabaticity in which the field is replaced directly with its time dependent form is all the story, because nonequilibrium behaviors such as violation of the fluctuation-dissipation theorem and even negative values of the susceptibility appear. Still, the scaling law among the critical exponents holds between the response and the correlation, either of which can thus be employed to estimate the critical exponents as in equilibrium.

We have identified a unique type of initial conditions that dominates the evolution under the driving. Opposite to the nonequilibrium initial conditions that lead to the critical initial slip, this type of initial conditions has longer correlations than the driving ones and can be in either equilibrium or nonequilibrium. An example of the latter is one arising from a driving that changes continuously to a subsequent one with a shorter dominant time scale. Under the latter driving, all correlations of the system are still governed remarkably by the time scales of the former driving in the initial-state-controlled regime.

Applications of the theory to some specific forms of driving have discovered results that have not been found before. Besides the negative susceptibility, a monomial driving with $n>1$ involves a singularity that originates from the difference in the scaling functions between their different forms of rescaled arguments and that are characterized completely by $n$. A polynomial driving can exhibit initial state dominated regimes when it crosses or approaches the critical point several times in addition to the different regimes and their crossovers determined by the coefficients. The other general driving is dominated by its first expansion coefficient in time, often the linear one, when its overall amplitude is sufficiently large. Some time scales determined by its parameters may only be transient or may be dominated by the initial state and thus do not control any regime before the initial state decays away. The presence of the amplitude involving rescaled arguments, no matter whether as a small variable in the scaling functions or more seriously as a variable that gives rise to crossovers, cautions experimental measurements in which an external driving is applied to a system with long relaxation times.

As the system studied is a generic nonlinear nonequilibrium one, the theory may shed light on the study of other nonequilibrium systems. It may also be instructive to nonlinear science as the driving may help to probe scaling behavior there.

\end{document}